\documentclass[showpacs,prb,twocolumn,groupedaddress,floatfix]{revtex4-1}
\usepackage{bm}
\usepackage{pstricks}
\usepackage{amsmath,amssymb,amsbsy}
\usepackage{graphicx,subfigure}
\usepackage{hyperref}
\usepackage{verbatim}
\usepackage[utf8]{inputenc}
\usepackage[USenglish]{babel}
\usepackage[T1]{fontenc}
\usepackage{graphicx}

\begin{document}
\title{Band nesting and the optical response of two-dimensional
semiconducting transition metal dichalcogenides}


\author{A. Carvalho$^{1}$, R.~M. Ribeiro$^{1,2}$, A.~H. Castro Neto$^{1}$}

\affiliation{$^{1}$Graphene Research Centre, National University of Singapore, 6
Science Drive 2, Singapore 117546}

\affiliation{$^{2}$Center of Physics  and  Department of Physics,
University of Minho, PT-4710-057, Braga, Portugal}

\date{\today}

\begin{abstract}
We have studied the optical conductivity of two-dimensional (2D) semiconducting transition metal dichalcogenides (STMDC) using {\it ab initio}
density functional theory (DFT). We find that this class of materials presents large optical response due to the phenomenon of
{\it band nesting}. The tendency towards band nesting is enhanced by the presence of van Hove singularities in the bandstructure of these
materials. Given that 2D crystals are atomically thin and naturally transparent, our results show that it is possible to have strong photon-electron
interactions even in 2D.
\end{abstract}
\maketitle

Semiconductor transition metal dichalcogenides (STMDC) are a family of crystals 
with a chemical formula {\cal M}{\cal X}$_2$ where
{\cal M} = W, Mo, Ti, Zr, Hf, Pd, Pt, and others,
and {\cal X} = S, Se, Te,\cite{wilson_charge-density_1975,wang_electronics_2012,chhowalla_chemistry_2013}
which can exist in a two-dimensional (2D) structure consisting of
 one layer of transition metal atoms sandwiched by two layers of chalcogens,
all in hexagonal sublattices.
They have two known structural polytypes,
trigonal prismatic (T) and octahedral (O),
which can be distinguished by the relative stacking of the chalcogenide layers.
Most 2D STMDC have band gaps in the visible
range, between 1 eV and 3 eV, and have been the subject of study in the last few years\cite{fuhrer2012,Mak2010}
since the emergence of the field of 2D crystals.\cite{neto_new_2011}
Because of these band gaps, in a technologically interesting range, these materials are being considered for a new generation
of 2D transistor, sensor, and photovoltaic applications.

It was discovered recently \cite{Britnell2013} that these materials have strong optical properties even when they are only three atoms thin.
This is rather surprising because atomically thin films like these, only tens of \AA ngstr\"{o}ms in thickness, are naturally transparent and we would not
expect a strong photon-electron coupling {\it a priori}. 
In this article, we show that this extraordinary optical response is due to the phenomenon
of ``band nesting", namely, the fact that in the bandstructure of these materials there are regions where conduction and valence bands are parallel
to each other in energy. Band nesting implies that when the material absorbs a photon, the produced electrons and holes propagate with exactly the
same, but opposite, velocities. 
We find that band nesting is present in the bandstructure of all these materials. 
Furthermore, the existence of strong van Hove singularities (VHS) facilitates the phenomenon of band nesting.
In two-dimensional materials, the band-nesting results in a divergence of the joint density of states,
leading to very high optical conductivity.
We present calculations of the optical response of the 2D STMDC with $X=$S,Se,
illustrating how it is enhanced by the phenomenon of band nesting.

\subsection{Band nesting}

In semiconductors, the band gap plays an important role in what concerns optical absorption. 
It defines the threshold after which there is absorption
of electromagnetic radiation, by the promotion of an electron from the valence band to the conduction band. 
But the largest absorption is usually not
at the band gap edge; 
it is often considered to be in a VHS in the electronic structure. 
These correspond to singularities in the density of states;
if at a given point of the reciprocal space there are VHS both in the conduction and the valence band, 
there will be a singularity of the optical conductivity. 
Yet, this coincidence normally happens only at high
symmetry points, 
and there are very few in the Brillouin Zone (BZ). A particular case is the extended van Hove singularity (EVHS) in that these are
single band saddle points with a flat band in one of the directions.\cite{gofron_observation_1994}

The optical conductivity of a material can be written as
\begin{equation}
 \sigma_1(\omega) = \kappa_2(\omega)\omega\epsilon_0 \, ,
\nonumber
\end{equation}
where $\kappa_2(\omega)$ is the imaginary part of the relative electric permittivity, $\omega$ is the frequency
of the incoming electromagnetic radiation, and $\epsilon_0$ is the vacuum permittivity. 
In the optical dipole approximation we can write:
\begin{equation}
\kappa_2(\omega) = A(\omega)\sum_{v,c}\int_{BZ} \dfrac{d^2\bf{k}}{(2\pi)^2}|d_{vc}|^2\delta\left( E_c - E_v - \hslash\omega \right) \, ,
\label{kappa2}
\end{equation}
The sum is over the occupied states in the valence band ($v$) and the unoccupied states in the conduction band ($c$)
with energies $E_v$ and $E_c$,
and includes implicitly the sum over spins, $A(\omega) = 4\pi^2e^2/(m^2\omega^2)$ ($e$ is the electric charge and $m$ the
carrier mass), $d_{vc}$ is the dipole matrix element. The integral in (\ref{kappa2}) is evaluated over the entire 2D BZ.
If we consider cuts $S(E)$ of constant energy $E$, $E = \hslash\omega = E_c - E_v$, in the bandstructure, we can write:
\begin{equation}
 d^2{\bf k} = dS\dfrac{d\left(E_c - E_v \right) }{|\nabla_k \left(E_c - E_v \right)|} \, ,
\nonumber
\end{equation}
and the integral in (\ref{kappa2}) can be rewritten as:
\begin{equation}
 \kappa_2(\omega) = A(\omega)\sum_{v,c} \dfrac{1}{(2\pi)^2}\int_{S(\omega)}\dfrac{dS}{|\nabla_k \left(E_c - E_v \right)|}|d_{vc}|^2 \, .
\nonumber
\end{equation}
Notice that the strong peaks in the optical conductivity will come from regions in the spectrum where
$|\nabla_k \left(E_c - E_v \right)|\approx 0$. If $d_{vc}$ varies slowly over these regions (so that there is a gradient expansion)
we can write:
\begin{equation}
 \kappa_2(\omega) \approx A(\omega)\sum_{v,c} |d_{vc}|^2 \rho_{vc}(\omega) \, ,
\nonumber
\end{equation}
where
\begin{equation}
 \rho_{vc}(\omega) = \dfrac{1}{(2\pi)^2}\int_{S(\omega)}\dfrac{dS}{|\nabla_k \left(E_c - E_v \right)|} \, ,
\nonumber
\end{equation}
is the joint density of states (JDOS).

The points where $\nabla_k \left(E_c - E_v \right) = 0$ are called critical points (CP) and they can be of several types.
If $\nabla_k E_c = \nabla_k E_v = 0$ we have either a maximum, a minimum or a saddle point in each band; 
this usually occurs only at high symmetry points. 
These points often receive more attention, because they are easy to pinpoint by visual
inspection of the bandstructure, and give rise to singularities in the DOS. On the other hand, the condition $\nabla_k \left(E_c - E_v \right) = 0$ with
$|\nabla_kE_c| \approx |\nabla_kE_v| > 0$, that is {\it band nesting}, 
gives rise to singularities of the JDOS, and therefore to high optical conductivity.
Notice that this condition differs from an EVHS \cite{gofron_observation_1994} in that the
later refers to saddle points in one band, with a flat band in one of the directions, 
while here it is determined by the ``topographic'' difference between the conduction  and valence bands.
In the case of two dimensional materials, a saddle point of $E_c-E_v$ gives rise to a divergence
of the optical conductivity, whereas in 3D materials it merely gives rise to an edge with $(E-E_0)^{1/2}$ dependence, in first approximation.\cite{bassani-book}

\section{Method}
We performed a series of DFT calculations for the STMDC family using the
open source code {\sc Quantum ESPRESSO}.\cite{Giannozzi2009}
We used norm conserving, fully relativistic pseudopotentials with nonlinear core-correction and spin-orbit information
to describe the ion cores.\cite{PSEUDO}
The exchange correlation energy was described by the generalized gradient approximation (GGA), in the scheme
proposed by Perdew, Burke and Ernzerhof\cite{Perdew1996} (PBE).
The integrations over the Brillouin-zone (BZ) were performed using scheme proposed by Monkhorst-Pack\cite{Monkhorst1968,dft}
for all calculations except those of the density of states, for which
the tetrahedron method\cite{Blochl1994} was used instead.
We calculated the optical conductivity directly from the bandstructure.\cite{epsFR}
It is well known that GGA underestimates the band gap,\cite{komsa-PRB-86-241201} 
and hence the optical conductivity shows the peaks displaced towards lower energies relative to actual
experiments. 
However, their shapes and intensities are expected to be correct.

We notice the importance of including spin-orbit and so to perform full relativistic, non-collinear calculations \cite{zhu_giant_2011,ramasubramaniam_tunable_2011}.
Significant spin-orbit splittings in the range 50~meV to 530~meV can be obtained in these crystals and can be measured using current spectroscopic techniques.
Still, spin-orbit interaction is ignored in most of DFT calculations \cite{ataca_stable_2012,bhattacharyya_semiconductor-metal_2012,ding_first_2011,kuc_influence_2011}. In our case, even for light transition metals, such as Ti, we can have a spin-orbit splitting of the order of 40~meV, which can be easily measured.
The trigonal prismatic (T) geometry does not have inversion symmetry, and has a considerable spin-orbit splitting, specially
around the high symmetry point K. 
The octahedral structure (O) has inversion symmetry, 
and therefore no spin-orbit splitting can be observed ($E(k,\uparrow)=E(k,\downarrow)$). 
This results from the inversion symmetry of the energy bands in the reciprocal space, 
which implies that $E(k,\uparrow)=E(-k,\uparrow)$ and $E(k,\downarrow)=E(-k,\downarrow)$, 
while time reversal symmetry (preservation of the Kramers degeneracy) 
requires that $E(k,\uparrow)=E(-k,\downarrow)$.

\section{Results}
\begin{figure*}[t]
\centering
 \includegraphics[width=1.4\columnwidth]{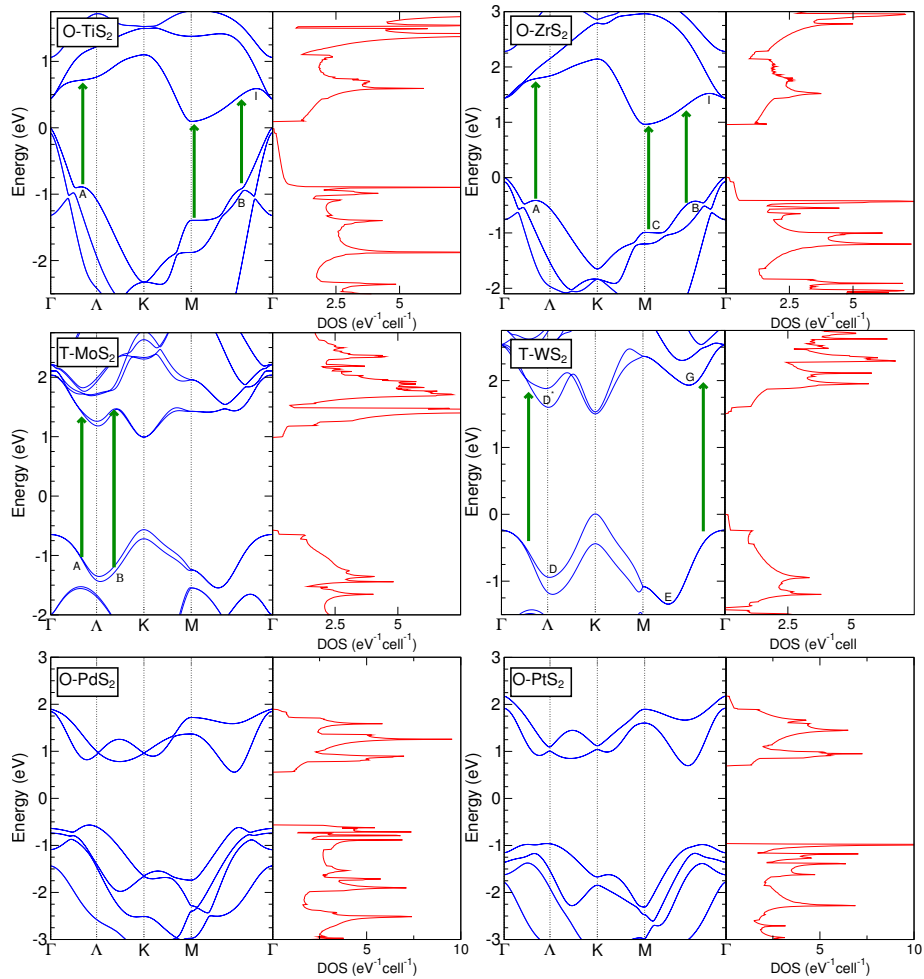}
 \caption{(Color online) Band structures, and DOS of TiS$_2$ and  ZrS$_2$ (group 4A sulphides), 
MoS$_2$ and WS$_2$ (group 6A sulphides) and PdS$_2$ and PtS$_2$ (group 8 sulphides).
 The arrows indicate the transitions corresponding to the first prominent peaks in the optical conductivity.}
 \label{fig:bands}
\end{figure*}

\subsection{Bandstructure calculations}

Calculations of the electronic structure were performed for all 2D $MX_2$ with $X=$S, Se,
for both the trigonal prismatic and octahedral structures.
Amongst these, we found eleven to be semiconductors.
Unless otherwise stated, we will only show results for the 
lowest energy structures for each compound, 
which are the T structure for Mo$X_2$ and W$X_2$ and the O structure for Ti$X_2$, Zr$X_2$, Pt$X_2$ and Pd$X_2$.
However, the same analysis can be extended to the metastable structures as well.

The electronic bandstructures and density of states (DOS)
of TiS$_2$, ZrS$_2$, MoS$_2$, WS$_2$, PtS$_2$ and PdS$_2$ are shown in Fig. \ref{fig:bands}.
It is useful to compare the results for dichalcogenides with $M$ 
belonging to the same group of the periodic table,
which usually have the same lowest energy structure type
and have similar features in the bandstructure close to the gap. 
The same can be said of $M$S$_2$ and $M$Se$_2$ for the same transition metal.
However, T and O structures, even of the same material, are very different.
Nevertheless, all of them present Van Hove singularities of $E_c$, $E_v$ or both, 
including saddle points which give rise to sharp peaks in the DOS.

\begin{figure}[h]
 \centering
 \includegraphics[keepaspectratio=true]{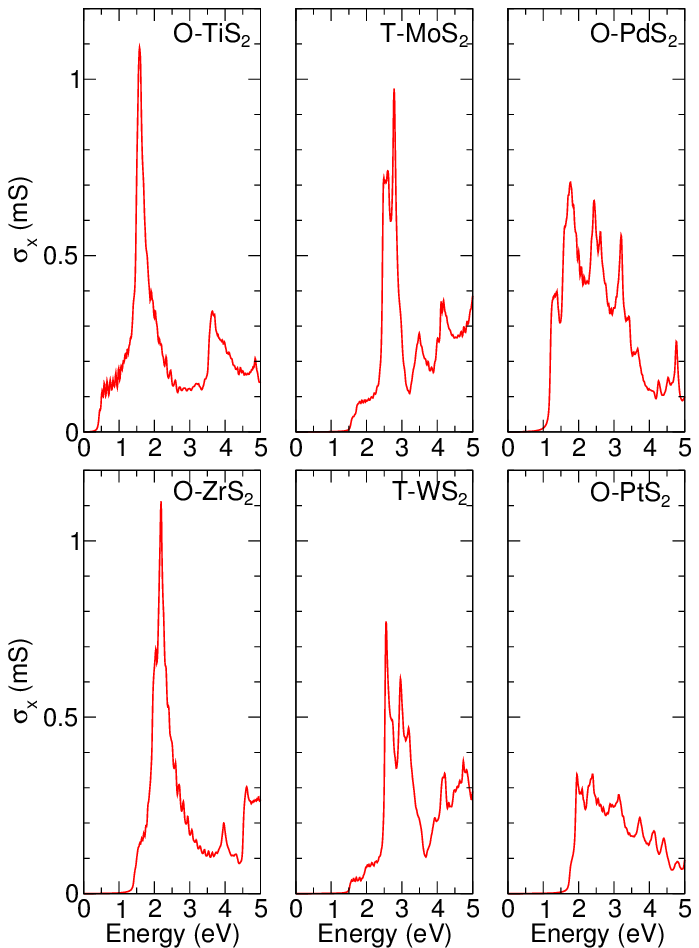}
 \caption{(Color online) Real part of the optical conductivity of 2D transition metal disulphides.}
 \label{fig:sigma}
\end{figure}

We start by analyzing the bandstructure of WS$_2$, one of the most studied STMDC.
At the K point,
where the direct gap is smallest,
the Van Hove singularities are the minimum of $E_c$ and maximum of $E_v$,
and therefore only give rise to steps of the DOS.
These steps are low compared to the sharp peaks originating on the very flat bands near
the conduction band minimum between the M and the $\Gamma$ points (see point marked as G in Fig. \ref{fig:bands}),
which is not a high symmetry point. 
Still, the singularity of the DOS itself is not sufficient to explain the high absorption peak that can be seen
in the optical conductivity (see Fig. \ref{fig:sigma}).

\begin{figure}[h]
 \centering
 \includegraphics{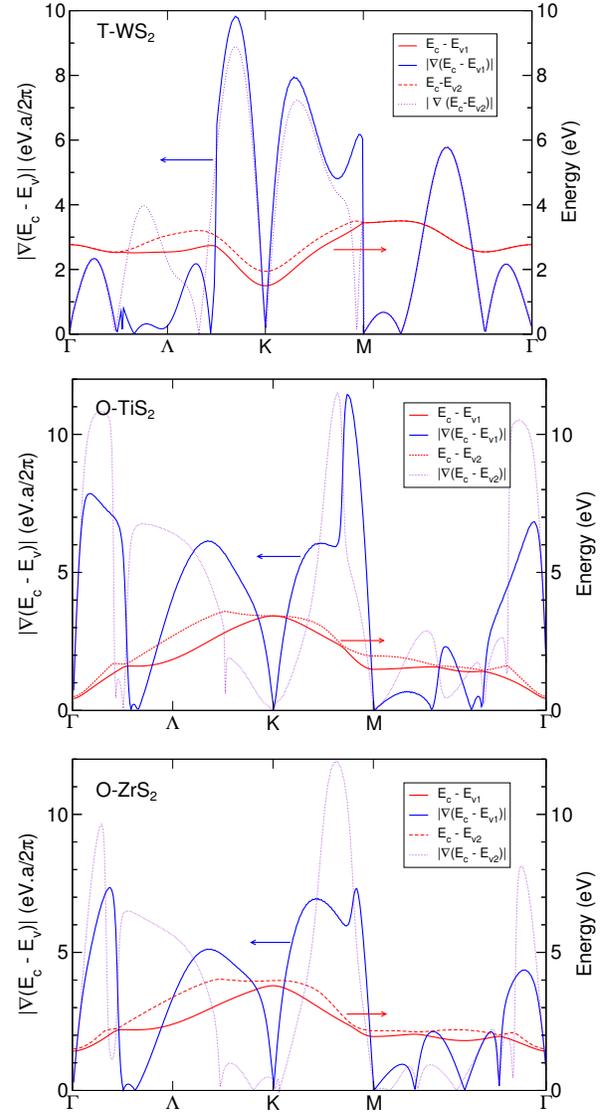}
 \caption{(Color online) Difference $E_c - E_v$ and the modulus of its gradient
 for monolayer WS$_2$, TiS$_2$ and ZrS$_2$
 in the high symmetry path.\cite{noteG}
 $E_{v1}$ indicates the highest occupied band, while $E_{v2}$ indicates the energy of the
second highest occupied band.
 $a$ is the lattice constant.}
 \label{fig:grad}
\end{figure}

In order to identify the origin of the largest peak at low energy (at 2.56 eV),
we analyze the energy difference between the 
lowest unoccupied band and the highest occupied band, 
$E_c-E_{v1}$ (the index of $E_c$ will be omitted for simplicity), 
together with its gradient, along the high symmetry lines of the Brillouin Zone
(Fig. \ref{fig:grad}).
We find the gradient to be very low between the $\Gamma$ and the $\Lambda$
points (corresponding to transitions signaled in Fig. \ref{fig:bands})
 which is the first large optical conductivity peak at 2.56~eV. 
It is also small near the right arrow of Fig. \ref{fig:bands}, at around 2.7~eV.
We define the regions where this band nesting occurs using the criteria
 $|\nabla_k \left(E_c - E_v \right)| \ll 1$~eV/($2\pi/a$)
(where $2\pi/a$ is the modulus of the reciprocal lattice vector).

\begin{figure}[h]
 \centering
 \includegraphics[scale=0.2,keepaspectratio=true]{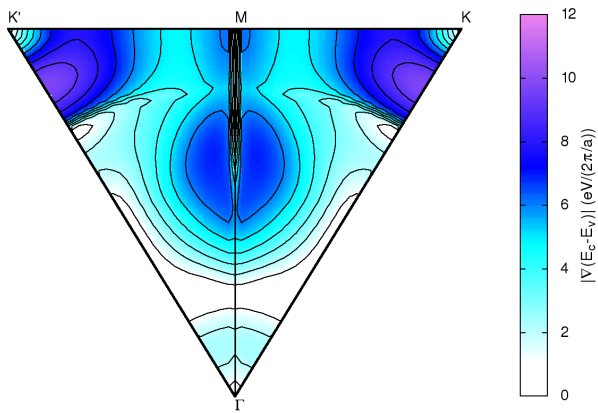}
 \caption{(Color online) Map on the BZ of $|\nabla_k \left(E_c - E_{v1} \right)|$ for WS$_2$.
 $a$ is the lattice constant.
In the $\Gamma$-M line, $\nabla_k \left(E_c - E_{v1} \right)$ is 
undefined due to band crossing.}
 \label{fig:map-Grad-T-WS2}
\end{figure}

We explored all the BZ to find the extent of the band nesting.
Figure \ref{fig:map-Grad-T-WS2} shows
 $|\nabla_k \left(E_c - E_{v1} \right)|$ for WS$_2$.
The large white areas close to $\Lambda$ are the areas 
where band nesting occurs for these two bands.

The band nesting can also be observed for other bands
immediately below or above, as for example for the transitions between the 
second highest band and the conduction band ($E_c-E_{v2}$),
also illustrated in Fig. \ref{fig:bands}.
For example the 2.96~eV peak in optical conductivity
results mostly from contributions of other bands.

The bandstructures of the other trigonal prismatic compounds,
WSe$_2$, MoS$_2$ and MoSe$_2$ display  similar  band nesting.

The band nesting is also present in the bandstructure of octahedral polytype compounds.
Figure \ref{fig:bands} shows the bandstructure and DOS of O-TiS$_2$ single layer.
This material exists in the bulk in the octahedral form, and was predicted to be an energetically
stable semi-metal \cite{ataca_stable_2012}. 
However, our calculations show it to be an indirect band gap semiconductor, with a small gap.
Experimentally, the bulk form of TiS$_2$ is a very narrow band gap semiconductor\cite{kukkonen_transport_1981,chen_angle-resolved_1980}
($E_g \approx 0.3$~eV). 
This value is probably underestimated due to the semilocal approximation
used for the exchange and correlation energy functional.
We also note that, since there is no spin-orbit splitting, all the
bands shown are degenerate, and so contribute doubly to the DOS.

Following the same reasoning we used for the trigonal prismatic materials and analyzing the energy gradients (Fig. \ref{fig:grad}),
we notice that $|\nabla_k \left(E_c - E_v \right)| \ll 1$~eV/($2\pi/a$) in the regions corresponding to the arrows of Fig. \ref{fig:bands}.
There is another band bellow, and very close in energy to the highest occupied band, 
which is also plotted in Fig. \ref{fig:bands}.
Since it has transition energies very close to the ones from the highest occupied band, it mostly reinforces the peaks
due to the band nesting. All the three transitions have similar energies, being the strongest near M at 1.5~eV; the others contribute to the large
broadening of the peak in the optical conductivity (Fig. \ref{fig:sigma}).

\begin{figure}[h]
 \centering
 \includegraphics[scale=0.2,keepaspectratio=true]{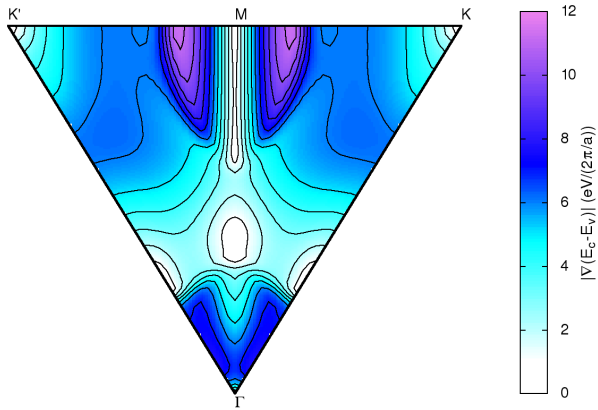}
 \caption{(Color online) Map on the BZ of $|\nabla_k \left(E_c - E_{v1} \right)|$ for TiS$_2$.
 $a$ is the lattice constant.}
 \label{fig:map-Grad-O-TiS2}
\end{figure}

We analyze the extent of this band nesting over the BZ by plotting 
$|\nabla_k \left(E_c - E_{v1} \right)|$ for TiS$_2$ (Figure \ref{fig:map-Grad-O-TiS2}).
In white we have the zone corresponding to values less than 1~eV/($2\pi/a$).
It can be seen that band nesting extends significantly beyond the high symmetry lines. The larger the area, the more intense the absorption peak is expected to be.

Another element of this family, ZrS$_2$, behaves in a similar way. ZrS$_2$ has the same octahedral structure
and the same number of valence electrons as TiS$_2$. But in this case, the gap is much wider (Fig. \ref{fig:bands}).

The transitions marked by the arrows in Fig. \ref{fig:bands}
correspond to regions where the gradient of $E_c-E_{v1}$ is small (Fig. \ref{fig:grad}).
Hence, the absorption is very high at these energies, as can be seen in Fig. \ref{fig:sigma}.
There we have two very close peaks, forming a very broad peak.
They correspond to a transition at the M point with an energy $E=2.0$~eV,
and the transition indicated by the letter A with an energy $E=2.2$~eV.
The transitions at B ($E=1.88$~eV) also give some contribution to the broadening of the peak in the optical conductivity.
The transition at M is even stronger than for TiS$_2$.
Both TiS$_2$ and ZrS$_2$ have absorption at lower energies than the corresponding to these transitions,
but the intensity is almost an order of magnitude smaller.
It is interesting to note that TiS$_2$ and ZrS$_2$ have a larger optical conductivity than the corresponding systems based
on W or Mo.

We have verified all these results for all elements of the 2D STMDC that include
WS$_2$, WSe$_2$, MoS$_2$, MoSe$_2$, in the trigonal form,
and TiS$_2$, ZrS$_2$, ZrSe$_2$, PdS$_2$, PdSe$_2$, PtS$_2$, PtSe$_2$ in the octahedral form
and the band nesting is qualitatively the same.
The only variation that we find is quantitative, namely, the intensity of the optical response changes from system to system (Fig. \ref{fig:sigma} shows that the high peaks near the absorption edge are
about half as high for PtS$_2$ and PdS$_2$ as for TiS$_2$, for example).
However, band nesting is present for all members of this family of 2D materials.

\section{Summary}
In conclusion, we have shown that all 2D STMDC display band nesting in large
regions of the Brillouin Zone.
This feature of their bandstructure leads to a large optical response and peaks in the optical conductivity.
The octahedral compounds TiS$_2$ and ZrS$_2$ are amongst those with largest band nesting regions.
The trigonal prismatic systems, which lack inversion symmetry, 
also have strong non-linear optical response.
This result indicates that despite their thickness, these materials present strong photon-electron coupling. 
The existence of large electron-photon interaction in 2D opens up the possibility to exciting
opportunities for basic research as well as for applications in photonics and opto-electronics.
\vspace*{0.2cm}


\begin{acknowledgments}
We gratefully acknowledge JJ Woo and MC Costa
and the computer resources from TACC and GRC.
RMR is thankful for the financial support by FEDER through the COMPETE
Program and by the Portuguese Foundation for Science and Technology
(FCT) in the framework of the Strategic Project PEST-C/FIS/UI607/2011
and grant nr. SFRH/BSAB/1249/2012. We acknowledge the NRF-CRP award
"Novel 2D materials with tailored properties: beyond graphene" (R-144-000-295-281).
\end{acknowledgments}



\begin{thebibliography}{99}
\bibitem{PSEUDO}%
  \BibitemOpen
  \bibinfo {note} {The pseudopotentials used were either obtained from the {\sc Quantum ESPRESSO}
  distribution or produced using the {\sc atomic} code by A. Dal Corso,
  that comes in the {\sc Quantum ESPRESSO} distribution.}%
  \bibAnnoteFile{Stop}{PSEUDO}%
\bibitem{dft}%
  \BibitemOpen
  \bibinfo {note} {Single layer samples were modeled in a slab geometry by
  including a vacuum region of 45~Bohr in the direction perpendicular to the
  surface. A grid of $16\times16\times1$~$\mathbf{k}$-points was used to sample
  the BZ. The energy cutoff was 50~Ry. The atomic positions were optimized
  using the Broyden-Fletcher-Goldfarb-Shanno (BFGS) method for the symmetric
  structure. The lattice parameter $a$ was determined by minimization of the
  total energy. The electronic density of states was calculated by sampling
  $48\times48\times1$ points of the BZ, and broadened with a 0.01~eV Gaussian
  width. A Gaussian broadening of 0.05~eV width was applied in the optical
  conductivity.}%
  \bibAnnoteFile{Stop}{dft}%
\bibitem{epsFR}%
  \BibitemOpen
  \bibinfo {note} {The dielectric permittivity and the optical conductivity
  were calculated using a modified version of the {\sc epsilon} program of the
  {\sc Quantum ESPRESSO} distribution, to account for full relativistic
  calculations.}%
  \bibAnnoteFile{Stop}{epsFR}%
\bibitem{noteG} For WS$_2$, $\left|\nabla_k (E_c-E_v)\right|$ is not defined in the line between M
and $\Gamma$. 
We plot instead the linear derivative along that line, $d/d\xi((E_c-E_v))$, 
where $\xi$ is a coordinate between M and $\Gamma$.
\end{thebibliography}

%
\end{document}